\documentclass[a4paper]{ESLAB2005}
\usepackage{epsfig}

\begin{document}

\title{Probing the birth of the first quasars with the future far infrared mission}

\author{M.\,J. Page%\inst{1} 
} \institute{Mullard Space Science Laboratory, University College London,
Holmbury St. Mary, Dorking, Surrey RH5 6NT, UK
}

\maketitle 

\begin{abstract}
\noindent
It is now widely recognised that massive black holes 
must have had a fundamental influence on the
formation of galaxies and vice versa. With current and imminent missions 
we aim to
unravel much of this relationship for the last 10 Gyr of cosmic history, 
while the quasar population
waned and star formation died down. The picture at earlier times will be more
difficult to reconstruct, but will likely be even more exciting: when the first
stars shone, the first dust was formed, and quasars were a vigorously rising
population. One of the primary goals of XEUS is to allow us to find and study
the earliest quasars, however deeply buried in gas and dust they may be. But to
understand fully the astrophysical context of these objects, their significance
in the grand picture, we must learn how they relate to their environments and
their host galaxies. The ESA future Far-InfraRed Mission (FIRM) will 
provide much of the
data we require, revealing the dust heated by star formation in the host
galaxy, the relative evolutionary stages of spheroid and black hole, and 
the total energy budgets posessed by these first quasars. FIRM
will reveal star formation in the immediate proto-cluster environment of
the quasar and so tell us how the formation of the first galaxies and quasars
coupled to the earliest large scale structures.
\end{abstract}

\section{Introduction}
\label{sec:intro}

Three hundred and eighty thousand years after the big bang, the first 
atoms formed, and the primordial background radiation scattered for 
the last time. As the radiation cooled, the Universe entered a dark 
age that would last for half a billion years. Eventually, the first 
stars, or the first quasars re-lit the Universe with optical and 
ultraviolet radiation. 

Whether the Universe was illuminated first by stars or by quasars is still 
unknown. However, what has become clear in the last decade is that these 
two sources 
of radiation are intimately linked within the galaxies that host them. 
Quasars are extremely luminous, compact sources of radiation found at the
centres of galaxies. They derive their power from the accretion of surrounding
material onto a massive 
($>10^{6}$~M$_{\odot}$) black hole.
Stars, on the other hand, are powered by nuclear fusion, 
e.g. of hydrogen into helium, in their cores.
The two processes, accretion and nuclear 
fusion, have been responsible for almost 
all of the energy that has been generated and radiated since the first 
atoms formed. 

In this paper, I will discuss the value of ESA's future Far InfraRed Mission
(FIRM) for understanding the relationship between stars and quasars. I will
begin by outlining our basic understanding of the evolution of the stellar and
black hole components of galaxies. 

\section{The evolution of black holes and stars}

There is now overwhelming evidence that the creation and fuelling of quasars is
related to galaxy formation. The quasar population has evolved strongly with
cosmic epoch, declining dramatically since its heyday at redshift $\sim 2$ 
(e.g. Page et~al. 1997, Croom et~al. 2004).
Data from a variety of sources suggest that star formation also peaked at
redshift 1--2 (Bunker et~al. 2004, 
Blain et~al. 1999 and references therein). 
These redshifts correspond to the epoch at which galaxies are expected to have
assembled according to the hierarchical cold dark matter cosmology 
(Kauffmann 1996). 
The discovery of massive dark objects, remnants of once-luminous
quasars in the bulges of many nearby galaxies, 
further demonstrates that the creation and fuelling of quasars is inextricably
linked to the formation of galaxies (Magorrian et~al. 1998).
Present day massive, quiescent black holes are found to have 
mass roughly proportional to that of the surrounding galaxy spheroid 
(Merritt \& Ferrarese 2001).

The simplest scenario one can imagine to produce such coupled stellar
and black hole components in present-day galaxies is for the stellar bulge
and the massive black hole to grow at the same time, from the same gas.
In this scenario we would expect quasars, which are black holes that
are growing quickly, to be hosted by stellar bulges that are rapidly
growing their stellar mass, and are therefore experiencing episodes
of prodigious star formation. However, recent observations show that the 
relationship between star formation and black hole growth is more complex 
than this. 
I will return to this question in Sections \ref{sec:qsos} and
\ref{sec:submmgals}. First, I will describe briefly the use of X-ray
observations to identify quasars, and the use of far infrared observations to 
measure star formation.

\section{Searching for quasars}
Material accreted by a quasar 
reaches a large velocity as it falls into the gravitational well of the 
black hole, and forms a disc where some of 
the kinetic energy is thermalised and radiated. 
Much of this energy emerges in the extreme ultraviolet region of the spectrum,
but it cannot be observed because of photoelectric absorption by interstellar
hydrogen in our Galaxy.
Close to the event 
horizon of the black hole, particles are accelerated to very high temperatures,
 and inverse-Compton scatter radiation from the disc, forming 
an X-ray emitting corona. This 
X-ray emission accounts for $\sim$10\% of the bolometric output of a 
typical quasar, a much larger fraction than is emitted in X-rays by 
normal stars and galaxies. Furthermore, as X-rays are very penetrating,
they allow us to find quasars even when they are quite heavily obscured by
gas and dust.

\begin{figure}[ht]
  \begin{center}
    \epsfig{file=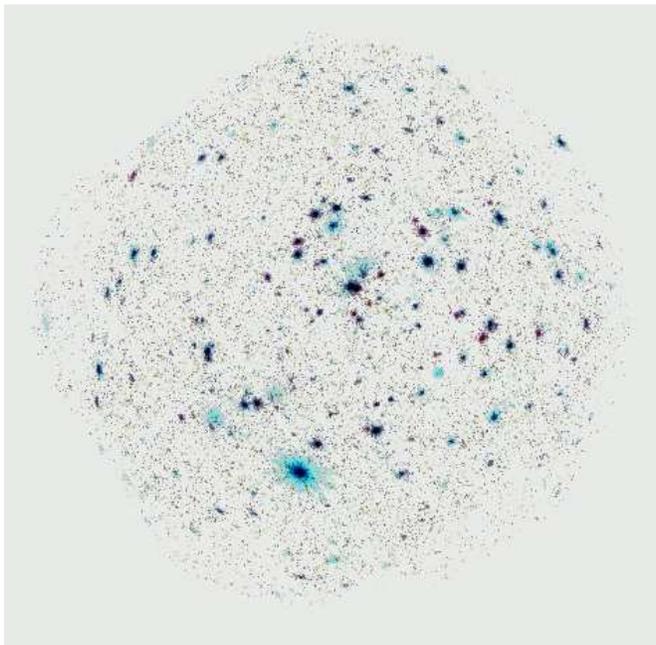, width=8.7cm}
  \end{center}
\caption{{\em XMM-Newton} image of the 13$^{H}$ deep field, a 30 arcminute
diameter region of sky
which has very little intervening Galactic material. Almost all the X-ray 
sources in the image are quasars.} 
\label{fig:13hr}
\end{figure}

Surveys of the X-ray sky therefore prove to
be an extremely good (probably the best) method of locating and identifying
quasars. It
turns out that the vast majority of sources that are detected in X-ray surveys
are quasars (e.g. Fig. \ref{fig:13hr}). 

\section{Searching for star formation} 
\begin{figure}[ht]
  \begin{center}
    \epsfig{file=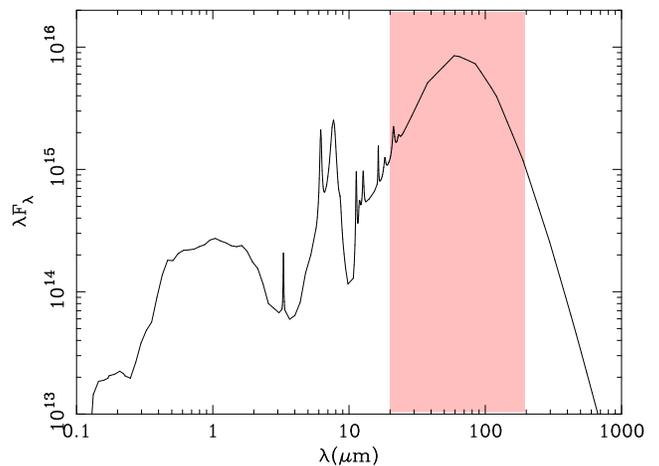, width=6.1cm, angle=270}
  \end{center}
\caption{Spectral energy distribution of the luminous starburst galaxy
Arp\,220. The spectrum is plotted as the product of wavelength ($\lambda$) and
the flux per unit wavelength ($F_{\lambda}$) so that constant energy per decade
in wavelength would be a horizontal line. 
The energy output is completely dominated by the thermal dust
emission between 20 and 200$\mu$m (shaded).} 
\label{fig:arp220sed}
\end{figure}

In star forming regions massive, short lived stars usually dominate the overall
radiative output. These stars emit most of their power in the rest-frame
ultraviolet, and so starburst galaxies are often identifiable by their strong
ultraviolet emission. However, the most vigorous starburst galaxies are often
so dusty that only a tiny fraction of their ultraviolet radiation manages to
escape. The spectral energy distribution of one such galaxy, Arp\,220, is shown
in Fig. \ref{fig:arp220sed}. In this case, almost all of the energy emerges in
the infrared part of the spectrum, between 20 and 200$\mu$m. It would be
impossible to determine the energy output of this galaxy without measurements
spanning the far infrared part of the spectrum. The spectral energy
distribution of Arp\,220 is thought to be fairly typical for the most powerful
starburst galaxies, and so surveys in the far infrared provide the most
reliable means of finding and identifying starburst galaxies.

\section{Star formation in quasar host galaxies}
\label{sec:qsos}

In total we expect the accreting black hole to 
emit about 1/5 as much 
energy as the stars forming in the surrounding galaxy spheroid. 
A far greater mass of 
gas is converted into stars, but accretion onto a black hole is a much 
more efficient means of producing radiation than nuclear fusion. Detecting 
the optical and ultraviolet radiation from the host galaxies of distant 
quasars is extremely challenging, because the quasar dominates the radiation in
these wavebands.
However, powerful star forming regions almost always contain large 
quantities of dust,
which absorb much of the starlight. Most of the energy is  re-emitted 
as thermal radiation in the far-infrared. Observations at long 
wavelengths can therefore reveal major bursts of star formation from their 
strong dust emission. In recent years some such observations have become
feasible from the ground, exploiting the atmospheric transmission 
windows between 350$\mu$m and 1.1mm.   

Pioneering millimetre and submillimetre observations of the most powerful 
quasars, at very high redshift ($z>3$), suggested that high star formation
rates 
may be an ubiquitous characteristic of quasar host galaxies (McMahon
et~al. 1994, Omont et al. 1996, Isaak et~al. 2003).
However, if we look at the distribution of quasar luminosities (their
`luminosity function') over a range of cosmic time, it is
straightforward to determine the luminosity range, and the period of
cosmic history, that contributed the most to present day black hole
mass. The luminosity function has a distinctive knee or break in its
shape at a luminosity which changes with cosmic time (Page et~al. 1997, Croom
et~al. 2004).  At luminosites
higher than this knee, quasars drop rapidly in numbers, so that the
most powerful objects are exceedingly rare compared to their lower
luminosity cousins. At luminosities lower than the knee, quasars
are more numerous, but not by a large amount. The
contribution of any part of the luminosity function to the growth of
black hole mass is proportional to product of the numbers and
luminosities of the objects. This product is a maximum at the knee of
the luminosity function, and $\sim 70\%$ of the instantaneous black
hole mass growth rate comes from $\pm 0.7$~dex of the break. 

The quasar population has changed dramatically with cosmic time, peaking in the
$1<z<3$ epoch, when the typical luminosity of a quasar was $\sim$20 times
larger than it is in the present day Universe.  Therefore in terms of
contribution to the present day mass density of black holes, the most important
quasars by far are those in the redshift interval $1<z<3$, with luminosities
around the break in the luminosity function.

\begin{figure}[ht]
  \begin{center}
    \epsfig{file=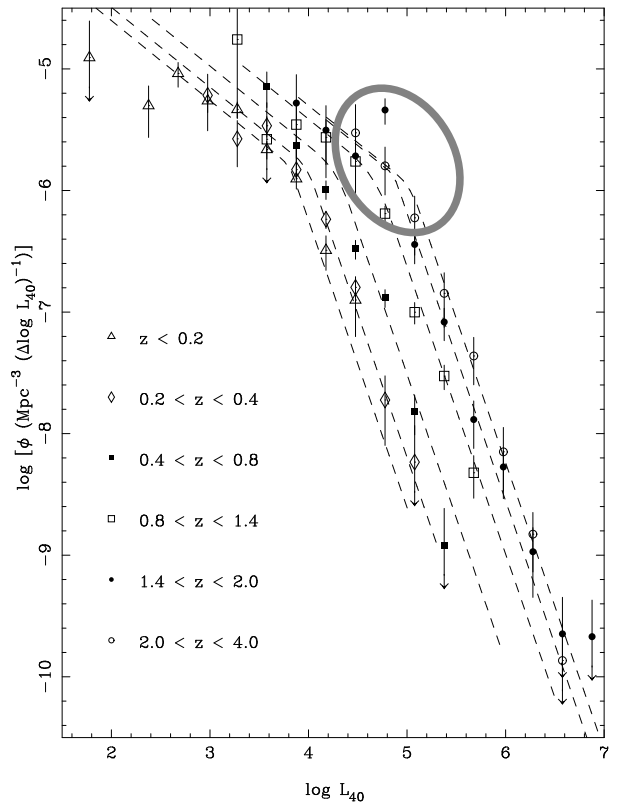, width=9.1cm}
  \end{center}
\caption{Quasar X-ray luminosity function (Page et~al. 1997). 
The grey ellipse
marks the knee in the
luminosity function at $z\sim2$, where the peak contribution to 
present day black
hole mass occurs.} 
\label{fig:lumifun}
\end{figure}

Submillimetre observations of quasars in this luminosity and redshift
range rule out the simple co-evolution models because
the quasars that are responsible for most of the black hole growth 
are undetectable
in the submillimetre. This means that they cannot be undergoing star
formation episodes of sufficient magnitude for the spheroid to build
up most of its mass in the same timespan as the black hole 
(Page et~al. 2004).

However, quasars at similar redshifts and luminosities, with normal quasar
spectra in the optical and UV, but with significant absorption in their X-ray 
spectra, are found to be luminous
submillimetre sources, undergoing major bursts of star
formation (Page et~al. 2001). 
This suggests an evolutionary sequence
in which X-ray absorbed quasars are at an earlier stage in their
evolution than the unabsorbed quasars, in line with a number of
theoretical models (Fabian 1999, Hopkins et~al. 2005). The
luminosities of the X-ray absorbed
quasars imply that they have already built up a significant fraction
of their ultimate black hole mass -- both the X-ray absorbed and
unabsorbed quasar phases are relatively late in the active quasars lifetime.
The relative numbers of X-ray absorbed and X-ray unabsorbed quasars suggest
that the X-ray absorbed phase is short, marking the transition from an earlier
heavily obscured phase to the emergence of a luminous, naked quasar.
When the quasar runs out of fuel, it ceases to shine, leaving an elliptical 
galaxy with a quiescent massive black hole in the centre.

\section{Black hole growth in starburst galaxies}
\label{sec:submmgals}

In the last 10 years, ground based surveys at submillimetre wavelengths have
revealed a remarkable population of 
 ultraluminous starburst galaxies, found
at high redshift (Smail, Ivison \& Blain 1997, Hughes et~al., 1998, Barger
et~al. 1998).
These objects are thought to
be massive galaxies undergoing their major episodes of star formation (Smail
et~al. 2002).
Recently, using the deepest X-ray observations ever taken, it has been 
possible to show that the
majority of these objects contain active accreting black holes
at their centres, often
obscured behind large column densities ($>10^{23}$~cm$^{-2}$) of gas and dust 
(Alexander et~al. 2005). The black holes
in these objects appear to be a factor of a few less massive and less luminous
than the quasars around the break in the luminosity function, placing them
earlier in the quasar evolutionary sequence. Nevertheless, with black holes of
$10^{7}$~M$_{\sun}$ and larger, these objects are already most of the way 
through their major black hole growth phases.

If we put together the observations of star formation in quasars, and of
black hole accretion in powerful starburst galaxies, we come to the following
picture:
\begin{itemize}
\item Submillimetre galaxies, X-ray absorbed quasars, 
X-ray unabsorbed quasars and elliptical galaxies appear to form an 
evolutionary sequence.
\item So far, we have only observed (or at least
recognised) the last $\sim 30\%$ of the quasar lifespan.
\item Absorption increases as we look further back in the sequence; the black
hole is likely to be very heavily obscured for the majority of its main
growth phase. 
\end{itemize}

\noindent
This picture has several implications for our way forward.  As we look
earlier in the history of a galaxy, the black hole gets more heavily buried in
gas and dust, so we need a more powerful X-ray telescope to be able to see
through the murk to the earlier stages of black hole growth. Most of the
accretion power will be absorbed by the surrounding gas and dust, and will be
reradiated in the infrared, but the galaxy will already be a bright source of
infrared emission because of the intense dust-enshrouded star-formation that
will be taking place. We can only disentangle the infrared quasar 
emission from the infrared starburst emission using spatial resolution or
detailed spectroscopy. Therefore we will need a far-infrared observatory with
very fine spatial resolution, excellent spectroscopic sensitivity, or both.
The Far InfraRed Mission (FIRM) identified in ESA's Cosmic Visions programme, 
fits the bill exactly.

\section{Quasars and the growth of structure}

In the currently favoured hierarchical cosmology, galaxy formation is a
consequence of the gravitational collapse of positive fluctuations 
in the large
scale density field. Small, galaxy-sized structures form first. 
Larger scale overdensities grow with time, drawing in
matter from their surroundings, ultimately producing the
filamentary ``soap bubble''  distribution seen in present day galaxies 
(Peacock et~al. 2001).
As the large scale structure develops, small galaxies merge to form 
larger galaxies, experiencing substantial bouts of star 
formation in the process; 
further mergers produce successively larger galaxies.
The most massive galaxies end up in the most overdense regions: those
which are destined to become clusters of galaxies by the present day.

\begin{figure}[ht]
  \begin{center}
    \epsfig{file=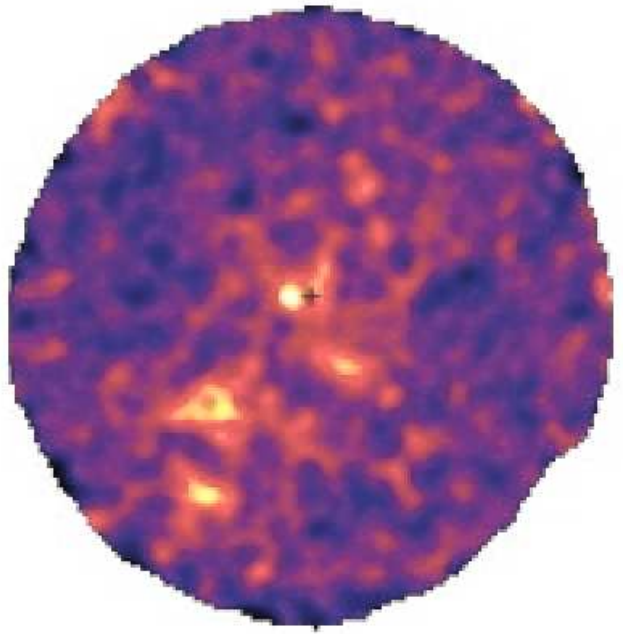, width=8.8cm}
  \end{center}
\caption{450$\mu$m image of a 2.5 arcminute region around the X-ray absorbed
quasar RXJ094144 (Stevens et~al. 2004), revealing a chain of ultraluminous 
starburst galaxies. The X-ray position of the quasar is marked by the cross 
in the centre of the image.} 
\label{fig:rxj0941}
\end{figure}

The black holes that once shone as powerful quasars now lie in the hearts of
massive elliptical galaxies, which in turn lie in clusters.  
By searching for merger-induced
starbursts within the environments of redshift $\sim 2$ quasars, we can examine
how the growth and evolution of massive black holes relate to the build up of
large scale structure. At present, with ground based observations, we can probe
only the most luminous starbursts in the most massive galaxies. A 450$\mu$m
image of the environment of the X-ray absorbed quasar RXJ094144 (Stevens
et~al. 2004) is shown in Fig. \ref{fig:rxj0941}. The image reveals a chain of
ultraluminous infrared galaxies around the quasar, with enough star formation
taking place for each starburst to evolve into a massive elliptical galaxy
within 1 Gyr.  In this case, the X-ray absorbed phase of the quasar coincides
with the formation of the massive cluster galaxies.  However, to measure the
dust emission and star formation rates of the many smaller galaxies that will
ultimately lie within the cluster will require much more sensitive
observations, at much higher spatial resolution.  Indeed most black holes 
lie in lower mass galaxies, and these within groups rather than
clusters of galaxies 
(our own Milky Way for example, lies within a group of galaxies). 
With FIRM, we will be able to probe energy production and star
formation in these numerous galaxies, and determine how the growth of 
black holes relates to the build up of structure in these smaller, 
more typical 
environments.

\section{A step into the far-IR with Herschel}
\begin{figure}[ht]
  \begin{center}
    \epsfig{file=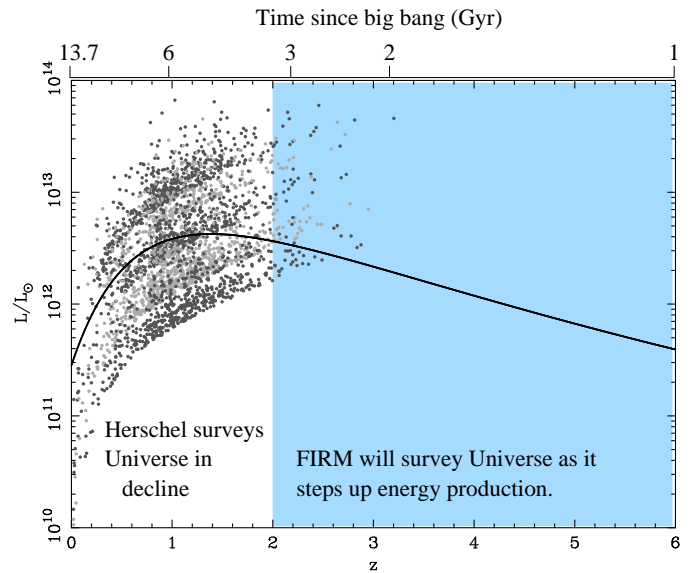, width=7.5cm, angle=270}
  \end{center}
\caption{Simulated coverage of the luminosity-redshift plane for a 5-tier
``wedding-cake'' {\em Herschel}-Spire survey at 250$\mu$m, similar to the 
surveys that are currently being planned. Each dot corresponds to an individual
galaxy; the individual layers of the cake (flux-limit/sky area combinations)
produce the 5 stripes of objects on the diagram. Cosmic time is 
indicated at the top of the plot. The solid curve shows the position of the
knee of the far infrared luminosity function as a function of redshift. On the
left hand side of the plot, where the Herschel surveys are effective, the
star formation density is declining with cosmic time. FIRM will be required to
probe the right hand side of the plot, where the energy output from star 
formation is still
increasing with cosmic time.} 
\label{fig:spiresurveys}
\end{figure}

The Herschel Space Observatory (Pilbratt 2003) 
will probe the Universe in the 60--600 $\mu$m
region. Due for launch in late 2007, with both spectroscopic and imaging
instruments, it will represent a huge advance in
this part of the spectrum.
With its deep extragalactic surveys, it is set to measure the star formation
for a large portion of cosmic history, and resolve a significant 
fraction of the 
far infrared
background into discrete sources.
Figure \ref{fig:spiresurveys} shows the anticipated coverage of a 5-tier
``wedding cake'' survey\footnote{A survey with five different sky-area/flux
limit combinations, designed to provide good sampling of redshift-luminosity 
space.} at 250$\mu$m. Herschel should perform extremely well in detecting
luminous infrared galaxies in the wavelength ranges where the bulk of their
bolometric power is emitted, out to redshifts of 2--3. This corresponds to a
large period of cosmic history ($>10$ Gyr), but for the majority of this period
star formation has been declining. Similarly, accretion onto massive black 
holes has been waning continuously since redshift 2; the powerful quasars are
past their prime, and each successive generation is less luminous than the 
last.

An earlier epoch, between 1 and 3 Gyrs after the big bang, is a much more
exciting period in the story of black hole growth. Star
formation was becoming more vigorous with time, and the massive black holes of
the most powerful quasars were growing exponentially, limited only by the
radiation pressure from their own central engines. From this period of 
black-hole gluttony emerged the massive compact objects that today lurk at
the centres of the greatest elliptical galaxies.
Unfortunately, it can be seen in Fig. \ref{fig:spiresurveys} that Herschel 
will only detect a small number of objects at the tail-end of this epoch. 
Principally, Herschel is limited by the angular resolution that is achievable
with its 3.5m primary mirror: source confusion
will make it impossible to detect the weaker, high 
redshift objects against the large sky density of brighter, foreground sources.
If we are to study the epoch of black hole growth, we will require 
a more sensitive 
far-infrared observatory, with much better angular resolution: FIRM.

\section{Chicken or egg at redshift 20?}

Which came first, stars or black holes? How did the first black holes come 
about?
Did the first stars become the first black holes, seeds around which whole
galaxies of stars would ultimately form?  These are arguably the most
fundamental of questions about massive black holes, and to answer them we will
have to make observations stretching right back into the dark ages of the
Universe, before reionisation. 

The initial results from the WMAP satellite suggest that reionisation occurred
at redshift $z=17\pm5$ (Bennett et~al. 2003), 
so the first stars and/or black holes must have formed
at $z\sim 20$. Not yet polluted by metals synthesised in stars, the 
primordial gas would have consisted almost
entirely of hydrogen and helium. The first collapsing gas clouds must
therefore have cooled primarily through molecular hydrogen (H$_{2}$) 
line emission. These emission lines are the key to identifying the first epoch
of star formation. The strongest lines predicted have rest frame wavelengths 
of 2--3 and 8--10$\mu$m (Mizusawa, Nishi \&
Omukai 2004, Ripamonti et~al. 2002, Kamaya \& Silk 2002). Although the highest
instantaneous luminosities are reached in the 2--3$\mu$m lines during the main
accretion phases of individual protostars, the 8--10$\mu$m 
lines are longer lived, and therefore more likely to be detected from an
assembly of star forming clouds. At redshifts of  15--20, the strongest 
lines will be
observed at 130--200$\mu$m: only with a facility 
such as FIRM can we hope to detect these lines and determine the time when the
first stars formed.

There are a whole variety of possibilities for the formation of the first black
holes. They could result from supernova explosions of the first generation of
stars, or they could form directly within primordial gas clouds. The latter
possibility requires the suppression of H$_{2}$ within the cloud, perhaps due
to UV radiation from the first stars (Bromm \& Loeb 2003). 
Such clouds could be identified by the
amount of atomic H\,I cooling relative to H$_{2}$ emission. FIRM observations 
will thus be key to timing and identifying the formation of the first black
holes relative to the formation of the first stars. 

\section{What do we need FIRM to be?}                                          

If we are to use FIRM to explore the origin, birth
and growth of massive black holes, we can identify the most basic requirements
as follows:\\

\noindent 
It must provide sensitive spectroscopy in the far infrared (25-300$\mu$m)
wavelength range.
\begin{itemize}
\vspace{-1mm}
\item It must be in space. 
\item It must have a cold aperture.
\end{itemize}
\noindent
It must have high enough spatial resolution that it is not confusion limited.
\begin{itemize}
\vspace{-1mm}
\item It could be a large ($>10$m) single dish.
\item It could be a multi-element interferometer.\\
\end{itemize}

\noindent
The most important decision to be taken, for the shape and capabilities of
FIRM, is whether to fly a single dish, or a multi-spacecraft 
interferometer.
At present both options are being
considered. A single dish would have superior surface brightness sensitivity,
but the interferometer wins out in spatial resolution. The two configurations
present different technical challenges, and these will have to be taken into
account along with the scientific trade-offs when the decision is made.

\section{XEUS and FIRM as partners}

Quasars are multiwavelength phenomena, emitting throughout the electromagnetic
spectrum, and our current understanding of them is the result of observations 
in every waveband. In the 2015--2025 timeframe, FIRM will be operating
alongside a number of exceptionally capable ground-based 
facilities covering a wide range of wavelengths, including the Atacama
Large Millimetre Array (ALMA), the Square Kilometre Array (SKA) and extremely
large (100m) optical/near-IR telescopes such as the European Southern
Observatory's OWL telescope. In addition, the gravitational wave observatory
LISA may be making a significant contribution to our understanding of black
hole growth via an entirely different form of radiation. 

However, for the study of the birth and growth of massive black holes, it is in
conjunction with ESA's next generation X-ray observatory XEUS that FIRM has the
greatest potential. The very large throughput of XEUS will enable it to detect
small, young quasars even when they are embedded in very dense cocoons of gas
and dust. Most of their radiation will be absorbed and re-emitted by the
surrounding material, and it is FIRM that will detect this radiation, so
telling us the total energy budgets of these quasars. FIRM will measure the
bolometric output from star formation in their host galaxies, telling us the
relative evolution of the black hole and the stellar components. The cryogenic
spectrometers on XEUS will identify outflows and winds from young quasars, that
may terminate the star formation by sweeping the cool gas from the host
galaxy. FIRM spectroscopy will provide the other half of the picture, by
revealing the mass, temperature, ionization state and dynamics of this cool 
gas.

As large scale structures developed, hot gas filled the potential
wells of clusters and groups that hosted powerful quasars.  XEUS will
detect this intracluster medium, and will allow us to measure the
conditions and elemental abundances as the gas built up. With FIRM we
will learn when, and how this relates to the star formation in the
galaxies of the cluster. FIRM will tell us the abundances and physical
conditions of cool gas within the galaxies so that we can follow the
enrichment history of the intracluster gas.  The combination of XEUS
and FIRM will allow us to determine the role of feedback from both
quasars and starbursts in galaxy formation, and in the heating of the
intergalactic medium.

\section{Conclusions}

In all directions X-ray telescopes reveal massive black holes at great
distances, in an earlier epoch, when they accreted material and shone as
quasars.  In the present day Universe, these black holes lie silent in
the centres of galaxies, with mass proportional to that of their
surrounding stellar spheroids.  This is most easily explained if the
formation of the two components was coeval, i.e. if the black hole was
built up by accretion of the same gas that rapidly formed the stars of
the spheroid. However, the picture is more complex observationally:
quasars which had redshifts and luminosities in the interval
responsible for most of today's black hole mass lived, for the most
part, in quiescent, finished host galaxies.  The formation of the
spheroid appeared to overlap the growth of the black hole only in
quasars which were hidden within cocoons of gas and dust, with
absorption increasing as we look to earlier times in the growth of the
black hole. Most of the black hole growth phase was probably heavily
obscured, suggesting that we have so far observed and recognised only
the final 30\% of the evolutionary sequence of a typical quasar.

To detect quasars in the earlier stages of their lives, we need a more
powerful X-ray telescope, XEUS, which can penetrate the dense gas and
dust in which they are buried. However to get the complete picture, we
will also require the Far InfraRed mission (FIRM). FIRM will have a
combination of sensitivity and spatial resolution that will allow it
to survey the Universe when the first galaxies were taking shape, when
quasars were still a vigorously rising population, and before star
formation reached its peak.  Where XEUS will detect the transmitted
radiation from youthful quasars, FIRM will measure the energy which
has been absorbed and re-emitted in the surrounding screens of gas and
dust; thus we will learn the total energy budgets of young quasars.
FIRM will measure the bolometric output from star formation
surrounding these quasars, to reveal how and when the obscured growth
of massive black holes took place relative to the build up of the
stars of their host galaxies. Finally, FIRM will detect star formation
in their immediate proto-cluster environments and thereby tell us how the
formation of the first galaxies and quasars coupled to the earliest
large scale structures.

\begin{acknowledgements}

Thanks to Tom Dwelly for providing Fig. \ref{fig:13hr}, and to Jason Stevens
for providing Fig. \ref{fig:rxj0941}.

\end{acknowledgements}

\end{document}